\documentclass[useAMS,usenatbib]{mn2e}

\usepackage{graphicx}
\usepackage{epsfig}

\usepackage{amsmath}  
\usepackage{tabularx}  
\usepackage{multirow}

\def\vhel{\ifmmode{V_{{\rm HEL}}}\else{$V_{{\rm HEL}}$}\fi}
\def\vsys{\ifmmode{V_{\rm sys}}\else{$V_{\rm sys}$}\fi}
\def\kms{\ifmmode{~{\rm km\,s}^{-1}}\else{~km~s$^{-1}$}\fi}
 \def\vlsr{\ifmmode{v_{\rm lsr}}\else{$v_{\rm lsr}$}\fi}

\def\ltsim{\ifmmode\stackrel{<}{_{\sim}}\else$\stackrel{<}{_{\sim}}$\fi}
\def\gtsim{\ifmmode\stackrel{>}{_{\sim}}\else$\stackrel{>}{_{\sim}}$\fi}

\def\reff@jnl#1{{\rm#1\/}}

\def\aj{\reff@jnl{AJ}}                  
\def\araa{\reff@jnl{ARA\&A}}            
\def\apj{\reff@jnl{ApJ}}                
\def\apjl{\reff@jnl{ApJ}}               
\def\apjs{\reff@jnl{ApJS}}              
\def\ao{\reff@jnl{Appl.Optics}}         
\def\apss{\reff@jnl{Ap\&SS}}            
\def\aap{\reff@jnl{A\&A}}               
\def\aapr{\reff@jnl{A\&A~Rev.}}         
\def\aaps{\reff@jnl{A\&AS}}             
\def\azh{\reff@jnl{AZh}}                        
\def\baas{\reff@jnl{BAAS}}              
\def\jrasc{\reff@jnl{JRASC}}            
\def\memras{\reff@jnl{MmRAS}}           
\def\mnras{\reff@jnl{MNRAS}}            
\def\pra{\reff@jnl{Phys.Rev.A}}         
\def\prb{\reff@jnl{Phys.Rev.B}}         
\def\prc{\reff@jnl{Phys.Rev.C}}         
\def\prd{\reff@jnl{Phys.Rev.D}}         
\def\prl{\reff@jnl{Phys.Rev.Lett}}      
\def\pasp{\reff@jnl{PASP}}              
\def\pasj{\reff@jnl{PASJ}}              
\def\qjras{\reff@jnl{QJRAS}}            
\def\skytel{\reff@jnl{S\&T}}            
\def\solphys{\reff@jnl{Solar~Phys.}}    
\def\sovast{\reff@jnl{Soviet~Ast.}}     
 \def\ssr{\reff@jnl{Space~Sci.Rev.}}     
\def\zap{\reff@jnl{ZAp}}                        
\def\nat{\reff@jnl{Nature}}             

\def\LaTeX{L\kern-.36em\raise.3ex\hbox{a}\kern-.15em
    T\kern-.1667em\lower.7ex\hbox{E}\kern-.125emX}

\begin{document}

\title[IR-correlated radio emission from Orion East]{IR-correlated 31~GHz radio emission from Orion East}

\author[Dickinson et al.]{C.~\!Dickinson,$\!^{1}$\thanks{E-mail: Clive.Dickinson@manchester.ac.uk} S.~\!Casassus,$\!^{2}$ R.~\!D.~\!Davies,$\!^{1}$ J.~\!R.~\!Allison,\!$^{3}$ R.~\!Bustos,$\!^{2,4}$ K.~\!Cleary,$\!^{5}$  
\newauthor R.~\!J.~\!Davis,$\!^{1}$ M.~\!E.~\!Jones,$\!^{3}$ T.~\!J.~\!Pearson,$\!^{5}$ A.~\!C.~\!S.~\!Readhead,$\!^{5}$ R.~\!Reeves,$\!^{5}$ 
\newauthor A.~\!C.~\!Taylor,$\!^{3}$ C.~\!T.~\!Tibbs,$\!^{1}$ R.~\!A.\!Watson$^{1}$ \\
$^1$Jodrell Bank Centre for Astrophysics, School of Physics \& Astronomy, University of Manchester, Oxford Road, Manchester, M13 9PL, U.K. \\ 
$^2$Departamento de Astronom{\'\i}a, Universidad de Chile, Casilla 36-D, Santiago, Chile\\
$^3$Oxford Astrophysics, University of Oxford, Denys Wilkinson Building, Keble Road, Oxford, OX1 3RH, U.K. \\
$^4$Departamento de Astronom{\'\i}a, Universidad de Concepci{\'o}n, Casilla 160-C, Concepci{\'o}ón, Chile \\
$^5$Cahill Center for Astronomy and Astrophysics, Mail Code 249-17, California Institute of Technology, Pasadena, CA 91125, U.S.A.
}


\date{Received **insert**; Accepted **insert**}
       
\pagerange{\pageref{firstpage}--\pageref{lastpage}} 
\pubyear{}

\maketitle
\label{firstpage}


\begin{abstract}
Lynds dark cloud LDN1622 represents one of the best examples of anomalous dust emission, possibly originating from small spinning dust grains. We present Cosmic Background Imager (CBI) 31~GHz data of LDN1621, a diffuse dark cloud to the north of LDN1622 in a region known as Orion East. A broken ring with diameter $\approx 20$~arcmin of diffuse emission is detected at 31~GHz, at $\approx 20-30$~mJy~beam$^{-1}$ with an angular resolution of $\approx 5$~arcmin. The ring-like structure is highly correlated with Far Infra-Red emission at $12-100~\mu$m with correlation coefficients of $r \approx 0.7-0.8$, significant at $\sim10\sigma$. Multi-frequency data are used to place constraints on other components of emission that could be contributing to the 31~GHz flux. An analysis of the GB6 survey maps at 4.85~GHz yields a $3\sigma$ upper limit on free-free emission of 7.2~mJy~beam$^{-1}$ ($\la 30$~per cent of the observed flux) at the CBI resolution. The bulk of the 31~GHz flux therefore appears to be mostly due to dust radiation. Aperture photometry, at an angular resolution of $13$~arcmin and with an aperture of diameter $30$~arcmin, allowed the use of IRAS maps and the {\it WMAP} 5-year W-band map at 93.5~GHz. A single modified blackbody model was fitted to the data to estimate the contribution from thermal dust, which amounts to $\sim 10$ per cent at 31~GHz. In this model, an excess of $1.52\pm 0.66$~Jy ($2.3\sigma$) is seen at 31~GHz. Future high frequency $\sim 100-1000$~GHz data, such as those from the {\it Planck} satellite, are required to accurately determine the thermal dust contribution at 31~GHz. Correlations with the IRAS $100~\mu$m gave a coupling coefficient of $18.1\pm4.4~\mu$K~(MJy/sr)$^{-1}$, consistent with the values found for LDN1622.

\end{abstract}



\setcounter{figure}{0} \normalsize

\section{INTRODUCTION}
\label{sec:introduction}

Anomalous Microwave Emission (AME) is the name given to excess microwave emission, observed at frequencies in the range $\sim10-60$~GHz, that is strongly correlated with far infrared (FIR) dust emission \citep{Leitch97,Kogut96,Finkbeiner02,Banday03,deOliveira-Costa04,Finkbeiner04,Watson05,Casassus08,Davies06,Dickinson09a,Scaife09,Tibbs10}. This dust-correlated emission is known to be a significant source of contamination for Cosmic Microwave Background (CMB) data, that must be separated accurately from the CMB signal \citep{Bennett03,Bonaldi07,Miville-Deschenes08,Gold09,Dickinson09b}. Although there is still some debate about the physical mechanism that is responsible for the emission, the most favoured explanation is in terms of small, rapidly spinning dust grains \citep{Draine98a,Draine98b}. Assuming this is the case, microwave observations of spinning dust represent a new way of studying the properties of interstellar dust grains and its environment within the interstellar medium \citep{Ali-Hamoud09,Dobler09,Ysard10}. Accurate observations, at frequencies in the range $5-100$~GHz, are now required to confirm spinning dust grains as the source of the anomalous emission and to study the spectrum to infer information about the grains and their environs.

One of the best examples of spinning dust emission, comes from observations of the dark cloud LDN1622 \citep{Finkbeiner02,Casassus06}. LDN1622 is a small ($\approx 10$~arcmin) dark cloud at the low Galactic latitude edge of the giant molecular cloud Orion B, on the outer edge of Barnard's loop \citep{Maddalena86}. The spectrum between 1~GHz and 3000~GHz is well-fitted by a superposition of optically thin free-free emission, thermal dust at a temperature of $T \sim 15$~K and a spinning dust component with a peak at $\sim 30$~GHz \citep{Casassus06}; at 30~GHz, and on angular scales $\la 20$~arcmin, the spectrum is dominated by spinning dust emission. Furthermore, correlations with IRAS infra-red maps indicate a better correlation with the emission from Very Small Grains (VSGs), as expected if the origin of the cm emission is spinning dust.

About $25$~arcmin to the north-east of LDN1622 lies another dark cloud, LDN1621, which is more diffuse than LDN1622 and forms a broken ring-like structure. These two clouds together were termed as {\it Orion East} by \cite{Herbig72} and are thought to lie at a distance of $\sim 400$~pc, although there is some debate about whether it lies much closer to us at $\sim140$~pc \citep{Wilson05,Kun08}. LDN1622 is a higher density dust cloud in which low mass star formation is beginning, as indicated by the presence of  T-Tauri stars \citep{Herbig72} but there are no bright OB stars in the vicinity; the UV radiation is predominately from the Orion OB1 association. By contrast, LDN1621 is a lower density, ring-like cloud, with no obvious pre-main sequence stars within \citep{Lee05}.

Here we present 31~GHz data of the area around LDN1621 and compare it with multi-frequency data, to estimate the contributions of free-free, thermal dust and anomalous dust components. This is compared with previous observations of the LDN1622 dark cloud to the south-west. Section~\ref{sec:data_reduction} describes the observations and data reduction while Section~\ref{sec:discussion} discusses multi-frequency maps of the LDN1621 region. Section~\ref{sec:analysis} gives a quantitative analysis of the contributions from free-free and thermal dust emissions and the possible origin of the 31~GHz excess. Conclusions are given in Section~\ref{sec:conclusions}.


\section{Observations and data reduction}
\label{sec:data_reduction}

The Cosmic Background Imager (CBI; \cite{Padin02}) is a 13-element close-packed interferometer, that operated at the Chajnantor Observatory, in the Atacama desert (Chile), from 1999-2008. The CBI was designed to image the CMB anisotropies on scales $\sim 5-30$~arcmin, in ten 1~GHz bands at $26-36$~GHz. With its high surface brightness sensitivity and calibration stability, it has also been used to image Galactic objects including diffuse dust clouds, supernova remnants, planetary nebulae and HII regions \citep{Casassus04,Casassus06,Casassus07,Casassus08,Hales04,Dickinson06,Dickinson07,Dickinson09a}. In 2006-2007, the CBI was fitted with 1.4~m dishes (CBI2), to allow optimal sensitivity on the smallest angular scales accessible by CBI (Taylor et al., in prep.). The larger dishes improved the point-source sensitivity and surface brightness sensitivity over angular scales $\sim 5-20$~arcmin. The primary beam is well-modelled by a Gaussian with a FWHM$=28.2$~arcmin at 31~GHz up to $\sim$FWHM/$2$, the region used in the analysis. 

During 2007 and 2008, CBI2 observed LDN1621 (J2000: R.A. $05^{\rm h}55^{\rm m}21.6^{\rm s}$ Dec. $+02^{\circ}11^{\rm m}33^{s}$) over 6 nights with a combined integration time of 5.7~hours on source. A trail field, separated by 8~min in R.A., was observed at the same hour angles to allow removal of local correlated signals such as ground spillover. The data analysis follows a similar procedure to that used in CMB data analysis (see \cite{Sievers09} and references therein). The amplitudes and phases were calibrated using short observations of Tau-A, assuming a flux density of 341~Jy at 31~GHz and a spectral index $\alpha=-0.299$ ($S \propto \nu^{\alpha}$). The absolute calibration is known to 0.5 per cent, with these values tied to the brightness temperature of Jupiter, assumed to be $T_{J}=146.6\pm0.75$~K \citep{Hill09}. All the data were inspected by eye to remove noisy data and baselines/antennas that were not providing adequate calibration. We assign an overall calibration uncertainty of $5$ per cent to account for residual gain variations, ground spillover and pointing errors.

\begin{figure}
\begin{center}
\includegraphics[width=0.46\textwidth,angle=0]{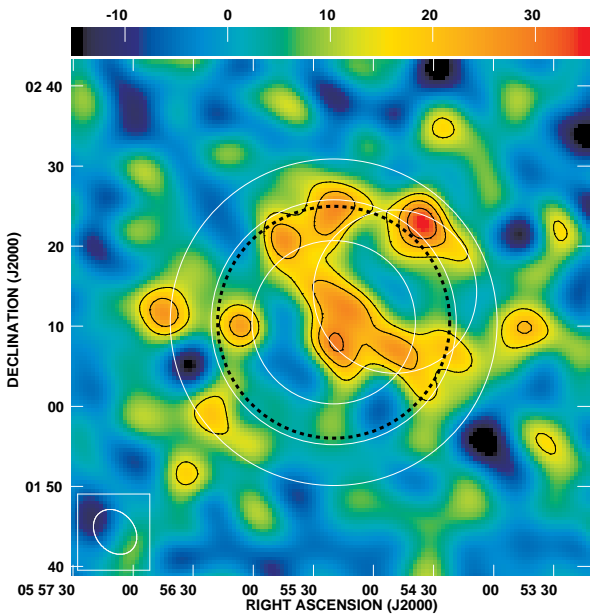}
 \caption{CLEANed 31~GHz map of the LDN1621 region. Contours are at 40, 60 and 80 per cent of the peak brightness (35.4~mJy~beam$^{-1}$). The synthesized beam ($6.1 \times 4.8$~arcmin) is shown in the lower-left corner. The FWHM of the primary beam is indicated by the thick black dashed line. Thin white solid lines indicate the apertures adopted in Section~\ref{sec:analysis}.  \label{fig:cbi_image}}
\end{center}
\end{figure}

The calibrated visibilities were inverted to form an image, and deconvolved using the CLEAN algorithm \citep{Hogbom74}, using the {\sc difmap} program \citep{Shepherd97}. Fig.~\ref{fig:cbi_image} shows the final CLEANed 31~GHz map of the LDN1621 region. We chose natural weighting of the visibilities to optimize for sensitivity rather than angular resolution. The CLEAN components were primary-beam corrected assuming a Gaussian model with FWHM$=28.2 (\nu_{\rm GHz}/31$)~arcmin. A low gain of 0.03 was used to recover as much extended emission as possible. The synthesized beam is $6.1\times4.8$~arcmin and the noise r.m.s. level in the map is $\approx 6$~mJy~beam$^{-1}$. Spectral indices within the CBI band were not reliable due to the extended nature of LDN1621 and limited signal-to-noise ratio with a small frequency lever arm of $10$~GHz.

\section{Discussion and comparison with multi-frequency data}
\label{sec:discussion}

The 31~GHz CBI image (Fig.~\ref{fig:cbi_image}) shows a ring of emission associated with the dark cloud LDN1621. The ring, of diameter $\approx 25$~arcmin, is not complete and is not of uniform brightness. The ring has a brightness in the range $\approx 20-30$~mJy~beam$^{-1}$, except for the northern part which is at $\approx 35$~mJy~beam$^{-1}$. It should be noted that the ring is not completely within the FWHM of the primary beam at 31~GHz, as indicated in Fig.~\ref{fig:cbi_image}. The northern part of the ring is at $\approx 15$~arcmin from the pointing centre and is therefore likely to be more affected by primary beam errors and noise. Indeed, the brightest pixel in the image is 16~arcmin from the pointing centre. We found that too many iterations of the CLEAN algorithm could result in additional flux in pixels outside the primary beam FWHM. In particular, for the brightest pixel, we found that the peak brightness could be boosted by up to a factor of $\approx 1.7$ compared to that presented here. Also, far outside the primary beam, very bright sources could be artificially produced. For this reason, bright pixels that were significantly outside of the primary beam FWHM were not corrected for the primary beam, to prevent distortion of the image. Inside the FWHM of the primary beam, such errors are negligible. We also verified the morphology of the 31~GHz image with an image created using earlier CBI1 data, of just 30 mins on a single night. Although of limited signal-to-noise ratio, the ring of LDN1621 was detected with a similar morphology to the data presented here, and the entire ring fits well within the primary beam FWHM of 45.2~arcmin at 31~GHz. It is also worth noting that LDN1621 is in fact visible on the edge of the CBI 31~GHz image of LDN1622 presented in \citet{Casassus06} and is seen to have a similar morphology.

\begin{figure*}
\begin{center}
\includegraphics[width=0.78\textwidth,angle=-90]{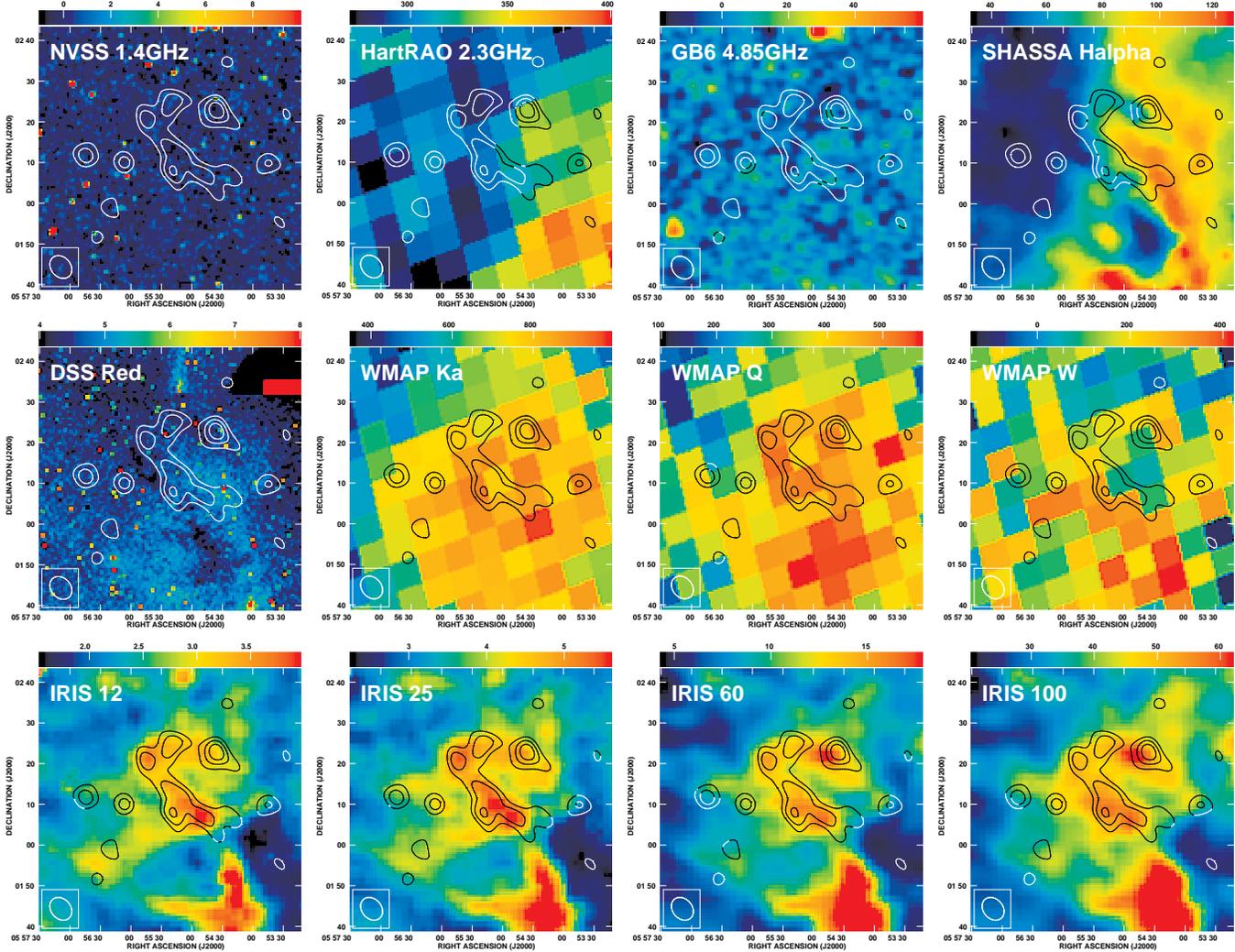}
 \caption{Multi-frequency maps of the LDN1621 region, superposed with the CBI 31~GHz contours from Fig.~\ref{fig:cbi_image}. The maps are, with colour scale units in parentheses, as follows: NVSS 1.4~GHz (mJy~beam$^{-1}$), HartRAO 2.3~GHz (mK), GB6 4.85~GHz (mJy~beam$^{-1}$), SHASSA continuum-subtracted H$\alpha$ (R), DSS red (arbitrary), WMAP Ka-band ($\mu$K), WMAP Q-band ($\mu$K), WMAP W-band ($\mu$K), IRIS $12~\mu$m (MJy/sr), IRIS $25~\mu$m (MJy/sr), IRIS $60~\mu$m (MJy/sr) and IRIS $100~\mu$m (MJy/sr). Each map is at the original angular resolution (see Table~\ref{tab:datasets}). \label{fig:multifreq_maps}}
\end{center}
\end{figure*}

Fig.~\ref{fig:multifreq_maps} shows a comparison of the 31~GHz contours (from Fig.~\ref{fig:cbi_image}) with various multi-frequency maps. The frequencies, angular resolutions and references are given in Table~\ref{tab:datasets}. The best IR data for comparison are the IRIS maps, which are the re-processed IRAS data at 12, 25, 60 and 100$~\mu$m \citep{Miville-Deschenes05}. We searched the {\it Spitzer} data archive for higher resolution IR data, but IRAC/MIPS data of the LDN1622 region only covered up to Dec.$\approx +02^{\circ}10^{\rm m}$ and therefore did not cover LDN1621. A ring of emission, coincident with the ring observed at 31~GHz, is clearly seen in the IRIS maps at $12-100~\mu$m. The IR ring of LDN1621 has a similar morphology to the 31~GHz contours. In particular, the ring is not complete to the west and has enhanced brightness on the north and south sides. The bright cloud to the south-west of the ring is the dark cloud LDN1622, and is not detected in these CBI data because of the primary beam attenuation beyond $\approx 20$~arcmin from the pointing centre.

The low frequency radio data at 1.4, 2.3 and 5~GHz (Fig.~\ref{fig:multifreq_maps}) do not show any evidence of emission associated with LDN1621. The relatively high angular resolution (45~arcsec) of the 1.4~GHz data is ideal for identifying extragalactic radio sources in the field that could be confusing the extended Galactic emission. No bright radio sources are detected in the field and no radio sources are coincident with the hot spots seen at 31~GHz down to a level of 2.5~mJy in the NVSS maps \citep{Condon98}. The majority of extragalactic sources have steeply falling spectra and therefore are unlikely to be detected at 31~GHz. Due to the limited $u,v$ coverage of the NVSS data, extended emission on scales of a few arcmin and larger will be resolved out by the interferometric response of the VLA, particularly in snap-shot mode. However, the 4.85~GHz GB6 image, at a resolution of 3.5~arcmin, also shows no evidence of emission in the vicinity at the level of $\ga8$~mJy~beam$^{-1}$. The GB6 maps are also affected by spatial filtering on scales $\ga 20$~arcmin \citep{Condon94} but remain sensitive to spatial scales comparable to the CBI. 

The {\it WMAP} 5-yr maps are the only other data available at frequencies close to 31~GHz. {\it WMAP} maps at Ka-band (33~GHz) and Q-band (40.7~GHz) show enhanced emission on large angular scales, while the ring is barely visible because of the limited angular resolution of the {\it WMAP} data (Table~\ref{tab:datasets}). At W-band (93.5~GHz), the large scale emission becomes faint and only LDN1622 is detected. In Section~\ref{sec:thermal_dust}, we use the W-band data to constrain the contribution from thermal dust.

The optical images of the LDN1621 region (Fig.~\ref{fig:multifreq_maps}) clearly show extended emission in the vicinity of LDN1621 and LDN1622. Most of this will be emission from the bright H$\alpha$ line, which has been separated in the SHASSA continuum-subtracted H$\alpha$ map. The H$\alpha$ line is a good tracer of warm ionized gas and because of its well-known dependence on the electron density ($I_{{\rm H}\alpha} \propto \int n_e^2 dl$), it is a good tracer of free-free emission \citep{Dickinson03}. There is a strong gradient of large-scale emission, becoming stronger to the south-west. This is expected since there is a strong UV-radiation field towards the Orion OB1 association in this direction. A large-scale gradient such as this will not be detected by the CBI as it is not sensitive to scales $\ga 20$~arcmin. An enhancement of H$\alpha$ intensity is visible around LDN1622 while it appears to be absorbing the majority of H$\alpha$ photons from LDN1622, which suggests that LDN1622 is in front of the ionized gas responsible for the H$\alpha$ emission. A similar situation occurs in the ring of LDN1621 but the effect is much less pronounced because the H$\alpha$ intensity is less and there is a stronger gradient at the position of LDN1621. Indeed, some estimates of the distance to LDN1621/1622 put it as close as 120~pc compared to 400~pc for Orion B \citep{Wilson05}.

\begin{table}
\caption{Ancillary data used in this paper. References are [1]: \citet{Condon98}; [2]: \citet{Jonas98}; [3]: \citet{Condon94}: [4]: \citet{Hinshaw09}; [5]: \citet{Miville-Deschenes05}; [6]: \citet{Gaustad01}; [7]: \citet{McLean00}.}
\begin{tabular}{lccc}
\hline
Dataset & Frequency/  & Beamwidth            &{Reference} \\
              &Wavelength   &(FWHM arcmin) & \\    \hline
NVSS          &1.4~GHz      &0.75                  & [1]  \\
HartRAO       &2.3~GHz      &20                    & [2]  \\
GB6           &4.85~GHz     &3.5                   & [3]  \\
WMAP Ka       &33.0~GHz     &$\approx 37$          & [4]  \\
WMAP Q        &40.7~GHz     &$\approx 29$          & [4]  \\
WMAP W        &93.5~GHz     &$\approx 13$          & [4]  \\
IRIS12        &$12~\mu$m    &3.8                   & [5]  \\
IRIS25        &$25~\mu$m    &3.8                   & [5]  \\
IRIS60        &$60~\mu$m    &4.0                   & [5]  \\
IRIS100       &$100~\mu$m   &4.3                   & [5]  \\
H$\alpha$     &656.2~nm     &$\approx 4$                   & [6] \\
DSS Red       &571~nm       &$\approx 0.03$        & [7] \\
\hline
\end{tabular}
\label{tab:datasets}
\end{table}

\section{Analysis \& Discussion}
\label{sec:analysis}

\subsection{Limits on free-free emission}
\label{sec:free-free}

The GB6 4.85~GHz data provide a strong upper limit on the contribution of free-free emission at 31~GHz. Steeper spectrum emission such as synchrotron radiation will be negligible at 31~GHz. No emission is detected within the LDN1621 area. After smoothing to the CBI resolution, the r.m.s. noise level at 4.85~GHz is $\approx 3$~mJy~beam$^{-1}$ corresponding to a $3\sigma$ upper limit of 9~mJy~beam$^{-1}$. At frequencies $\ga1$~GHz, free-free emission will be optically thin, except for the densest clouds, such as ultracompact HII regions. We can therefore reliably extrapolate the upper limit to 31~GHz assuming a flux density spectral index ($S \propto \nu^{\alpha}$) of $\alpha=-0.12$, appropriate for optically thin free-free emission \citep{Dickinson03}. The $3\sigma$ upper limit to free-free emission at 31~GHz is then 7.2~mJy~beam$^{-1}$. We can therefore be quite confident that the bulk of the 31~GHz emission cannot be due to synchrotron or free-free emission, although a small (up to a maximum of $\sim 30$ per cent) contribution from free-free contribution cannot be ruled out with the data available.

The level of free-free emission can also be estimated from H$\alpha$ data. If we assume that all of the absorbing dust lies in front of LDN1621, we can estimate the true H$\alpha$ intensity by estimating the dust extinction along this line of sight. Along the LDN1621 ring, the total H$\alpha$ intensity is $I_{{\rm H}\alpha} \simeq 40-70$~R. This is below the background value of $\sim100$~R and therefore the CBI will only detect a small fraction of this on scales $\approx 4-20$~arcmin. The dust absorption can be estimated using the column-density map, $D^{\rm T}$ at $100~\mu$m, of \citet{Schlegel98} assuming typical ISM conditions and reddening values. If we assume all the absorbing dust is in front of the H$\alpha$-emitting gas, the corrected H$\alpha$ intensity is given by $I_{{\rm H}\alpha}^{\rm corr}=I_{{\rm H}\alpha} \times 10^{0.0185D^{\rm T}}$ \citep{Dickinson03}, where $D^{\rm T}$ is the temperature-corrected intensity map in units of MJy/sr. Along the ring, $D^{\rm T} \simeq 50-60$~MJy/sr, thus $I_{{\rm H}\alpha}^{\rm corr} \sim 500$~R. For a typical electron temperature of $7000$~K, we expect a 31~GHz brightness temperature of $5.44~\mu$K/R \citep{Dickinson03}, or 2.7~mK at 31~GHz. This corresponds to $\sim 200$~mJy~beam$^{-1}$ at 31~GHz at the CBI resolution. However, this estimate is based on assuming all the dust along the line-of-sight is absorbing and that all the H$\alpha$-correlated emission is detected by the CBI. We can see from the SHASSA image that the bulk of H$\alpha$ emission is on large-scales and thus would not be detected by the CBI. The fluctuations in H$\alpha$ on scales of $\sim 5$~arcmin are at the $10-20$~R level, which corresponds to $\sim 10-20$~mJy~beam$^{-1}$ at 31~GHz, if all the dust is in front of the H$\alpha$-emitting gas. 

We must therefore rely on the low frequency data for the strongest constraints on free-free emission on these angular scales. We also note that there is a filament of H$\alpha$ that enters the the LDN1621 ring at a level of $\simeq 100$~R (see Fig.~\ref{fig:multifreq_maps}) but is not detected by the CBI. Given the noise level in the 31~GHz image is $\simeq 6$~mJy~beam$^{-1}$, this could be considered an approximate upper limit on the the free-free contribution, which is clearly small on CBI angular scales.

\subsection{Limits on vibrational (thermal) dust emission}
\label{sec:thermal_dust}

The 31~GHz data are highly correlated with the IR data at $12-100~\mu$m. The IR data are dominated by thermal emission from dust grains heated by UV radiation. The emission is often approximated by a modified blackbody function, $I_{\rm d} \propto \nu^{\beta+1}B(\nu,T_{\rm d})$, defined by an emissivity index, $\beta$, and a blackbody dust temperature, $T_{\rm d}$. With typical ISM dust temperatures in the range $T_{\rm d}\sim 10-50$~K, there is a peak at $\sim 3000$~GHz ($100~\mu$m). Typically, the Rayleigh-Jeans tail of this emission dominates down to $\sim 100$~GHz and becomes negligible at lower frequencies where synchrotron and free-free emission dominate. For dust grains in the ISM, an emissivity index of $\beta \sim 2$ is typical, although a range of values has been observed, reflecting different grain properties and size populations (see \citet{Dupac03} and references therein). It is therefore conceivable that the CBI 31~GHz data could simply be the long wavelength tail of the thermal dust. 

With no high resolution IR data available below $3000$~GHz, it is somewhat difficult to extrapolate accurately the thermal dust from IRAS data alone; we would need to assume a fixed value for the emissivity index. However, the highest frequency channel of {\it WMAP} data \citep{Hinshaw09} at W-band (93.5~GHz) has a resolution of $\approx 13$~arcmin and will be useful for placing limits on the level of the  thermal dust contribution, at least on these angular scales. Fig.~\ref{fig:multifreq_maps} shows the {\it WMAP} 5-year 93.5~GHz map. There is no strong emission detected at the location of LDN1621, although LDN1622 is visible to the south-west and there is low level emission just to the south of LDN1621. 

To make a more quantitative estimate of the thermal dust contribution, we smoothed the IRIS maps to a common resolution of 13~arcmin to allow a comparison with the 93.5~GHz data. We then measured the flux density in an aperture of diameter $30$~arcmin centred on LDN1621 (see Fig.~\ref{fig:cbi_image}). An estimate of the background level was estimated from an equivalent aperture placed in the top-right hand corner of the image (R.A. $05^{\rm h}53^{\rm m}14^{\rm s}$ Dec. $+02^{\circ}33^{\rm m}18^{\rm s}$) and subtracted from the LDN1621 aperture. Errors were derived from the r.m.s. fluctuations in the map. The results are given in Table~\ref{tab:aperture_photometry}. To obtain an equivalent 31~GHz value, we need to account for flux that has been lost due to the incomplete $u,v$-coverage of the CBI observations. We therefore estimated the flux density based on the map at $100~\mu$m, scaled to 31~GHz via the correlation coefficient derived in Section~\ref{sec:anomalous_emission} of $18.1\pm4.0~\mu$K~(MJy/sr)$^{-1}$; note that this is equivalent to calculating the amount of flux loss based on the $100~\mu$m morphology and correcting the CBI observed integrated flux density. The aperture analysis yielded an integrated flux density of $1.67\pm0.66$~Jy at 31~GHz. The error includes the r.m.s. from the sky and the error in the correlation coefficient.

\begin{figure}
\begin{center}
\includegraphics[width=0.45\textwidth,angle=0]{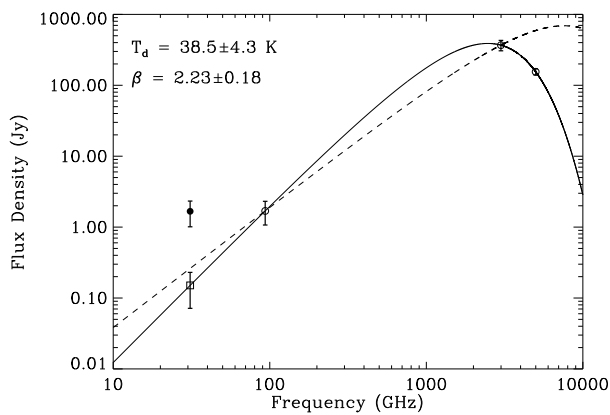}
 \caption{Spectrum of LDN1621 for 13~arcmin resolution data, within a 30~arcmin aperture (see text). The solid line represents the best-fitting modified blackbody curve using the 93.5, 2997 and 4995~GHz data points. The predicted 31~GHz flux density is shown as an open square. The actual observed value (estimated via cross-correlation with the $100~\mu$m map; see text) is shown as a filled circle. The dashed line is for a fit where the emissivity index is fixed at $\beta=+1.7$. \label{fig:sed}}
\end{center}
\end{figure}

The spectrum based on the aperture analysis is shown in Fig.~\ref{fig:sed}. We fitted a modified blackbody curve to the 93.5, 2997 and 4995~GHz data points only. The higher frequency data points ($12/25~\mu$m) do not give a good fit to such a simple model because the grains are not in thermal equilibrium with the radiation field and there are emission lines (ionic and PAH) within the $12/25~\mu$m passbands of IRAS. We do not consider the free-free component here because the GB6 4.85~GHz data are not reliable when integrating over regions $\ga 20$~arcmin.  The best-fitting model provides a good fit to the data with $T_{\rm d}=38.5\pm4.3$~K and $\beta=2.23\pm 0.18$. The dust temperature is somewhat higher than expected for a dark cloud; we may be seeing the effects of the local diffuse ISM in the region which may be warmer than the dust grains truly associated with LDN1621. The coupling coefficient is close to the expected value, and is constrained mostly by the W-band data point. We also tried fixing the coupling coefficient and/or emissivity index, to see if an adequate fit could be found but with a flatter index. However, it is not possible to find a different best fit to all three data points with just a single dust component. An example is shown in Fig.~\ref{fig:sed} where the emissivity index was fixed at $\beta=+1.7$; the best-fit is clearly not consistent with the data and the excess at 31~GHz still remains. However, we do note that multiple dust components could, in principle, still contribute at 31~GHz. An example would be a colder $(T_{\rm d}\la 10$~K) component, yet it cannot be dominant since the total emission must remain consistent with the 93.5~GHz data point. High sensitivity, high resolution data in the range $\sim 100-1000$~GHz, such as those to come from {\it Planck} \citep{Tauber10}, are required to rule this out and to obtain precise values for the thermal dust contribution.

\begin{table}
\caption{Integrated flux densities for a 30~arcmin diameter aperture centred on LDN1621, at 13~arcmin resolution. $^{*}$Note that the 31~GHz value was estimated by simulating the 31~GHz map based on the cross-correlation with the $100~\mu$m template.}
\begin{center}
\begin{tabular}{cc}
\hline
Frequency   &Flux density  \\
(GHz)       &(Jy)          \\
\hline
$31^{*}$          &$1.67\pm0.66$ \\
93.5        &$1.69\pm0.62$ \\
2997        &$368\pm62$    \\
4995        &$155\pm16$    \\
11988       &$47.8\pm3.3$  \\
24975       &$26.5\pm2.3$  \\
\hline
\end{tabular}
\end{center}
\label{tab:aperture_photometry}
\end{table}

\subsection{Anomalous microwave emission}
\label{sec:anomalous_emission}

In Sections~\ref{sec:free-free} and \ref{sec:thermal_dust} we used multi-frequency data to place limits on free-free and thermal dust emission in LDN1621 at 31~GHz, as observed by the CBI. The bulk of the emission at 31~GHz appears to be anomalous in that it cannot be easily explained by the usual diffuse components of the ISM at radio/microwave frequencies. Extrapolating the best-fitting thermal dust model curve to 31~GHz (Fig.~\ref{fig:sed}), gives a predicted 31~GHz flux density of $0.151\pm 0.079$~Jy. This is well below the estimated true value at 31~GHz of $1.67\pm0.66$~Jy. There is therefore an excess of $1.52\pm0.66$~Jy, which is significant at the $2.3\sigma$ level. For a fixed emissivity index, $\beta=+1.5$, the significance reduces to $1.8\sigma$. For an extreme value of $\beta=+1.0$, the flattest value that is expected on theoretical grounds, the significance is just $0.3\sigma$. High frequency data ($\nu \ga 100$~GHz), such as those expected from the {\it Planck} satellite, will be crucial in determining the contribution from thermal dust emission.

A number of possibilities exist to explain the excess 31~GHz emission including electro-dipole radiation from spinning dust grains \citep{Draine98a,Draine98b,Ali-Hamoud09,Ysard10}, magneto-dipole radiation \citep{Draine99}, free-free from hot $T\sim10^6$~K gas \citep{Leitch97}, flat-spectrum (hard) synchrotron \citep{Bennett03} and low-level solid-state structural transitions \citep{Jones09}. With the current radio/IR data, it is somewhat difficult to confidently distinguish between these possibilities. In other clouds, and in the diffuse ISM, the most favoured explanation is in terms of spinning dust grains. 

Several authors have used correlation analyses against dust templates to try to isolate the type of dust grains that correlate best with the radio data (e.g. \citealt{deOliveira-Costa02,Casassus06,Scaife09,Ysard10}). It is clear from Fig.~\ref{fig:multifreq_maps} that there is a good correlation with the IRIS maps at $12-100~\mu$m, suggesting a possible dust origin. We calculated Pearson's correlation coefficient, $r$, between the CBI 31~GHz map and IRIS data at 12, 25, 60 and 100~$\mu$m. We take into account the spatial filtering of the CBI instrument by simulating CBI observations, based on the real $u,v$ data with IRIS maps as inputs, and making CLEANed maps in the same way as for the CBI data. Errors were estimated by using Fisher's $r$ to $z$ transformation and calculating the $68$~per cent confidence interval \cite{Fisher15}. Since there is some concern regarding the primary beam correction beyond FWHM/2, we computed the correlation for different sized regions. Table~\ref{tab:correlations} gives the $r$ correlation coefficients for four regions. Given the noise level, and uncertainty in the primary beam correction in the outer regions, the 31~GHz-IR correlation is remarkably high with $r\sim 0.6-0.8$; significant at $\sim 10\sigma$. For spinning dust, we would expect a better correlation with shorter IR wavelengths ($\sim 10-30~\mu$m) because these are dominated by the smaller (and generally warmer) dust grains which are responsible for the bulk of spinning dust emission \citep{Ali-Hamoud09,Ysard10}. Longer wavelengths ($\ga 100~\mu$m) are dominated by larger (and typically cooler) grains. We find no strong preference for a particular IR template or short versus long IR wavelengths. The exception is for the 10~arcmin region centred in the middle of the ring, where there is a slightly better correlation with short ($12/25~\mu$m) wavelengths ($r=0.82\pm0.06$) compared with longer ($60/100~\mu$m) wavelengths ($r=0.69\pm0.09$). This is consistent with the expectation for spinning dust.

A number of authors have calculated the coupling coefficient between IR templates and the radio data. The coefficient for $100~\mu$m, sometimes referred to as the emissivity, is $\sim 10~\mu$K~(MJy/sr)$^{-1}$ in the diffuse ISM with a scatter of a factor of $\sim 2.5$ \citep{Banday03,Davies06}. In LDN1622 the coupling coefficient is $21.3 \pm 0.6~\mu$K~(MJy/sr)$^{-1}$ while HII regions are typically at $\la 5~\mu$K~(MJy/sr)$^{-1}$ \citep{Dickinson06,Dickinson07,Dickinson09a,Scaife08}. In LDN1621, taking no account of any free-free contribution, the coupling coefficient is $16-22~\mu$K~(MJy/sr)$^{-1}$ depending on which part of the image is used. For a 10~arcmin radius circle centred in the middle of the ring (see Fig.~\ref{fig:cbi_image}), where the spatial correlation is highest (Table~\ref{tab:correlations}), the coefficient is $18.1\pm4.0~\mu$K~(MJy/sr)$^{-1}$. Assuming there is no free-free contribution, which would reduce the coefficient, LDN1621 appears to emit at the same level as LDN1622 compared to the $100~\mu$m template, and a factor of $\approx 2$ more than the average value found in the diffuse ISM.

\begin{table}
\caption{Pearson correlation coefficients between the 31~GHz data and IRIS maps at 12, 25, 60 and 100$~\mu$m. These were calculated over circular regions with radii given in the first column. The regions are centred on the middle of the image except for the aperture marked with a asterisk ($^{*}$) which was centred in the middle of the ring feature at R.A. $05^{\rm h}54^{\rm m}46^{\rm s}$, Dec. $+02^{\circ}14^{\rm m}03^{\rm s}$ (see Fig.~\ref{fig:cbi_image}).} 
\begin{tabular}{lcccc}
\hline
Radius            &$12~\mu$m    &$25~\mu$m     &$60~\mu$m    &$100~\mu$m  \\
(arcmin)          &             &              &             &            \\
\hline
$10^{*}$ &$0.78^{\pm0.09}$ &$0.85^{\pm0.07}$ &$0.72^{\pm0.11}$ &$0.62^{\pm0.15}$ \\
10  &$0.75^{\pm0.11}$ &$0.70^{\pm0.12}$ &$0.75^{\pm0.11}$ &$0.71^{\pm0.12}$ \\
15  &$0.58^{\pm0.10}$ &$0.63^{\pm0.09}$ &$0.66^{\pm0.09}$ &$0.57^{\pm0.10}$ \\
20  &$0.63^{\pm0.07}$ &$0.68^{\pm0.07}$ &$0.70^{\pm0.06}$ &$0.62^{\pm0.08}$ \\
\hline
\end{tabular}
\label{tab:correlations}
\end{table}


\section{Conclusions}
\label{sec:conclusions}

LDN1621 is a region of diffuse emission $\approx 25$~arcmin to the north of LDN1622. Observations with the CBI at 31~GHz show a broken ring of emission, that is strongly correlated with FIR emission at $12-100~\mu$m, with Pearson correlation coefficients in the range $\approx 0.6-0.8$. Optical and H$\alpha$ data show absorption of a strong background of emission from warm ionized gas in the Eastern arm of Orion. This suggests that LDN1621 and LDN1622 are in the foreground of Orion (at a distance of $\sim 500$~parsec), possibly as close as 120~pc \citep{Wilson05}. No H$\alpha$ emission, associated directly with LDN1621, is seen. This suggests that LDN1621 itself is not emitting significant free-free emission, although the effects of dust extinction do not allow a strong constraint to be placed. Low frequency radio data also do not show evidence of diffuse emission associated with LDN1621. The 31~GHz emission is at $\approx 20-30$~mJy~beam$^{-1}$ while an analysis of the  GB6 map at 4.85~GHz provides a strong ($3\sigma$) upper limit of 7.2~mJy~beam$^{-1}$ at 31~GHz for free-free emission. The FIR-correlated emission at 31~GHz therefore appears to be mostly due to radiation associated with dust.

IRAS data alone do not allow a reliable extrapolation of the Rayleigh-Jeans thermal dust tail to 31~GHz. {\it WMAP} data at 93.5~GHz combined with IRAS data allowed the flux density to be estimated in an aperture of diameter 30~arcmin at an angular resolution of 13~arcmin. A single modified blackbody indicates that the thermal dust is $\sim 10$~per cent of the total 31~GHz flux, corresponding to an excess of $1.52\pm0.66$~Jy ($2.3\sigma$). The dust-correlated emission has a coupling coefficient, relative to $100~\mu$m, of $18.1\pm4.4~\mu$K~(MJy/sr)$^{-1}$, consistent with that observed from LDN1622. 

Orion East (consisting of both LDN1621 and LDN1622) appear to be part of the same system of dust clouds, emitting significant anomalous emission at frequencies $\sim 30$~GHz. Spinning dust is an obvious candidate for the physical mechanism responsible for the bulk of the emission. High sensitivity data, covering a wide range of frequencies ($\sim5-300$~GHz), is required to study such clouds in more detail. Data from the {\it Planck} satellite will be particularly useful in constraining the Rayleigh-Jeans dust tail, which may be responsible for a significant fraction of the 31~GHz if the emissivity index flattens at longer wavelengths.


\section*{Acknowledgments}

This work was supported by the Strategic Alliance for the Implementation of New Technologies (SAINT - see www.astro.caltech.edu/chajnantor/saint/index.html) and we are most grateful to the SAINT partners for their strong support. We gratefully acknowledge support from the Kavli
  Operating Institute and thank B. Rawn and S. Rawn Jr. The CBI was supported by NSF grants 9802989, 0098734
  and 0206416, and a Royal Society Small Research Grant. We are
  particularly indebted to the engineers who maintained and operated the
  CBI: Crist{\'o}bal Achermann, Jos{\'e} Cort{\'e}s, Crist{\'o}bal Jara, Nolberto Oyarace, Martin Shepherd and Carlos Verdugo. CD acknowledges an STFC Advanced Fellowship and ERC grant under the FP7. We acknowledge the use of the Legacy Archive for Microwave Background Data Analysis (LAMBDA). Support for LAMBDA is provided by the NASA Office of Space Science. We used data from the Southern H-Alpha Sky Survey Atlas (SHASSA), which is supported by the National Science Foundation.


\bibliographystyle{mn2e}

\bsp 
\label{lastpage}
\end{document}